\documentclass[twocolumn,aps]{revtex4-2}
\usepackage{graphicx}
\usepackage{bm}
\usepackage[linktocpage=true,colorlinks=true,urlcolor=blue,linkcolor=red,citecolor = blue]{hyperref}
\usepackage{pgfplots}
\usepackage{amsfonts}
\usepackage{amsmath}
\usepackage{amssymb}
\usepackage{xcolor}
\usepackage{braket}
\usepackage{relsize}
\usetikzlibrary{arrows}
\usepackage{enumerate}
\usepackage{pgfplots}
\usepackage{newfloat}
\usepackage{siunitx}
\usepackage{physics}
\usepackage{mathtools}
\usepackage{comment}
\usepackage{soul}
\usepackage[T1]{fontenc}

\DeclareFloatingEnvironment[name={Supplementary Figure}]{suppfigure}

\graphicspath{ {Figures/} }
\begin{document}

\title{Coupling conduction-band valleys in SiGe heterostructures via shear strain and Ge concentration oscillations}
\author{Benjamin D. Woods}
\email{bwoods6@wisc.edu}
\author{Hudaiba Soomro}
\author{E. S. Joseph}
\author{Collin C. D. Frink}
\author{Robert Joynt}
\author{M. A. Eriksson}
\author{Mark Friesen}
\affiliation{Department of Physics, University of Wisconsin-Madison, Madison, WI 53706, USA}

%TC:ignore
\begin{abstract}
    Engineering conduction-band valley couplings is a key challenge for Si-based spin qubits. Recent work has shown that the most reliable method for enhancing valley couplings entails adding Ge concentration oscillations to the quantum well. However, ultrashort oscillation periods are difficult to grow, while long oscillation periods do not provide useful improvements. Here, we show that the main benefits of short-wavelength oscillations can be achieved in long-wavelength structures through a second-order coupling process involving Brillouin-zone folding induced by shear strain. We finally show that such strain can be achieved through common fabrication techniques, making this an exceptionally promising system for scalable quantum computing.
\end{abstract}

\maketitle
%}
%TC:endignore

\section{Introduction}
Si/SiGe quantum dots are an attractive platform for quantum computation \cite{Loss1998,Hanson2007,Burkard2023}, as demonstrated by recent experiments \cite{Xue2022,Noiri2022,Mills2021} where both single and two-qubit gate fidelities have exceeded the error correction threshold of $99\%$ \cite{Fowler2012}. However, Si/SiGe quantum dots suffer from small energy spacings between the ground and excited valley states \cite{Zwanenburg2013}, called valley splittings, $E_\text{VS}$, which induce leakage if they are small~\cite{Buterakos2021}. Popular strategies for enhancing $E_\text{VS}$ include engineering atomically sharp interfaces~ \cite{Boykin2004a,Boykin2004b,Friesen2007} or narrow quantum wells~\cite{Losert2023,Chen2021}.
Unfortunately, random-alloy disorder~\cite{Losert2023} causes $E_\text{VS}$ to vary significantly from dot to dot~\cite{Chen2021,Wuetz2021}, with typical values ranging from $20$ to $300~\mu\text{eV}$ \cite{Borselli2011,Zajac2015,Mi2017,Scarlino2017,Neyens2018,Mi2018,Borjans2019,Hollmann2020,Dodson2022}.

The Wiggle Well (WW) has recently emerged as an important tool for enhancing the valley splitting, by adding Ge concentration oscillations (``wiggles") of wavelength $\lambda$ to the Si quantum well~\cite{McJunkin2021}.
Theoretical estimates suggest that these wiggles could provide a remarkable boost of $E_{\text{VS}} \approx 1~\text{meV}$ for an average Ge concentration of only $\bar{n}_\text{Ge} = 1\%$~\cite{Feng2022}.
However, the small wavelength needed for this structure ($\lambda =0.32~\text{nm}$) corresponds to roughly two atomic monolayers, suggesting that such growth would be very challenging.  
Naive band-structure estimates also identify a second, longer $\lambda$, that could produce a large $E_{\text{VS}}$ by coupling valleys in neighboring Brillouin zones.
However, rigorous calculations~\cite{Feng2022} show that this coupling is inhibited by the nonsymmorphic screw symmetry of the Si crystal~\cite{Woods2023a}.

In this work, we show that a modified structure called a strain-assisted Wiggle Well (STRAWW) can overcome these problems.
This device combines a \emph{long-wavelength} WW ($\lambda \approx 1.7~\text{nm}$), which has already been demonstrated in the laboratory~\cite{McJunkin2021}, with an experimentally feasible level of shear strain. 
We further show that valley splitting in STRAWW is largely independent of both the vertical electric field and the interface sharpness, thus simplifying heterostructure growth. 
We also note that the long-wavelength WW has a much larger spin-orbit coupling than conventional Si/SiGe quantum wells~\cite{Woods2023a}, even in the absence of shear strain, which enables fast spin manipulation via electric dipole spin resonance (EDSR), without requiring a micromagnet. This combination of large valley splitting and large spin-orbit coupling makes STRAWW a very attractive platform for Si spin qubits.
Below, we first provide an overview of the physics and implementation of strain in the STRAWW proposal, and then present calculations showing deterministically enhanced valley splittings.

%%%%%%%%%%%%%%%%%%%%%%%%%%%%%%%%
%%%%%%%%%%%%%%%%%%%%%%%%%%%%
\begin{figure*}[t]
\begin{center}
\includegraphics[width=.98\textwidth]{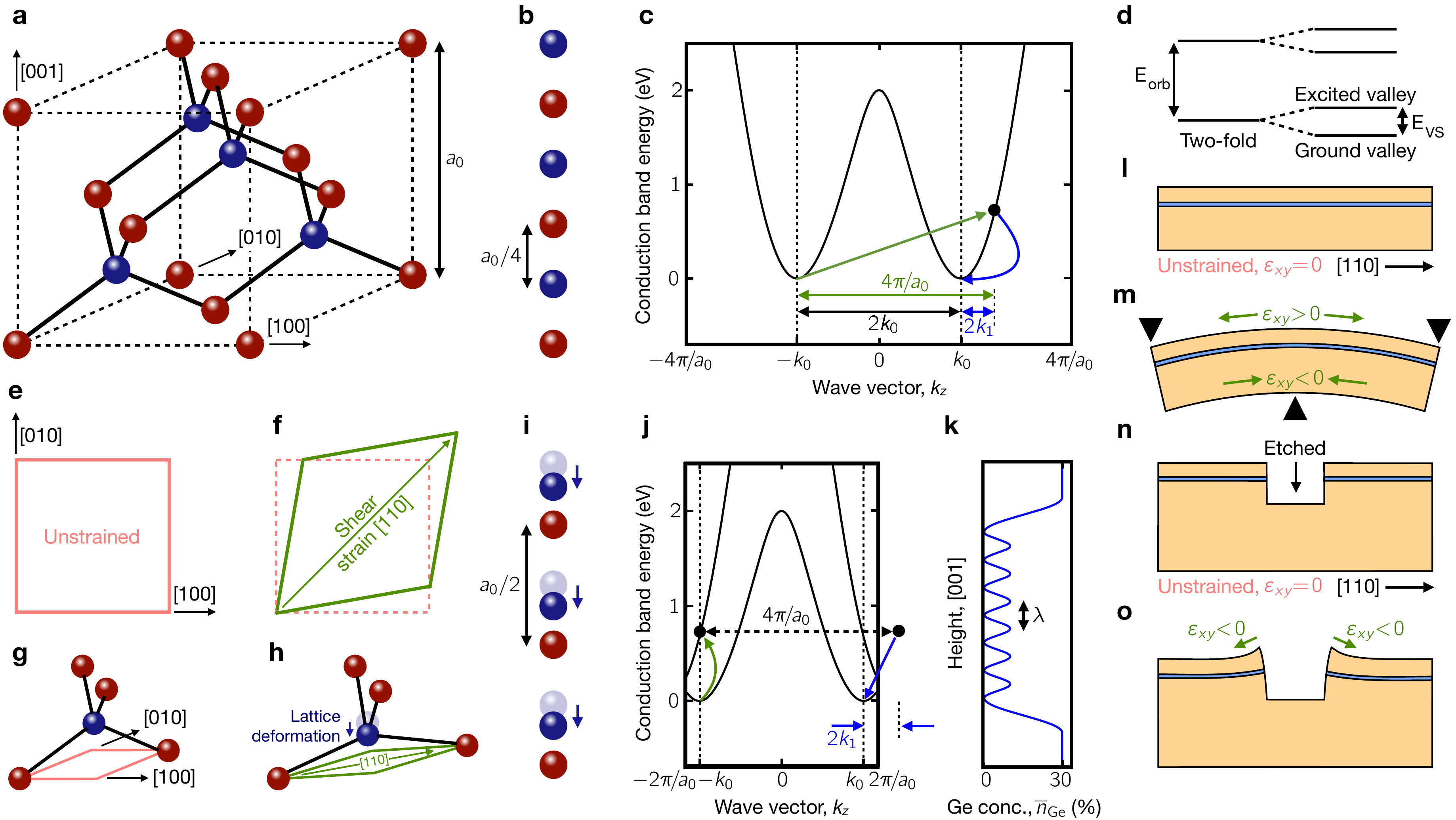}
\end{center}
\vspace{-0.5cm}
\caption{\textbf{Strain-assisted Wiggle Well (STRAWW) proposal.} 
\textbf{a}, Unstrained silicon cubic unit cell of length $a_0$, composed of two sublattices (red and blue). In the absence of strain, the sublattices are interchanged by a screw-symmetry operation. \textbf{b}, Assuming periodicity in the crystallographic [100] and [010] directions, the three dimensional (3D) lattice can be reduced to an effective 1D lattice \cite{Woods2023a}, whose primitive unit cell length is given by $a_0/4$, due to the screw symmetry. \textbf{c}, Low-energy Si band structure for the 1D lattice, expressed in a Brillouin zone (BZ) of length $8\pi/a_0$ along the [001] reciprocal axis $k_z$. 
The $2k_0$ valley coupling is achieved in two steps: (i) a $4\pi/a_0$ coupling induced by shear strain; (ii) a $2k_1$ coupling provided by the Wiggle Well. 
\textbf{d}, In the absence of valley coupling (left side), the states are two-fold degenerate due to the two-fold valley degeneracy in \textbf{c}. Valley coupling breaks this degeneracy (right side), leading to a valley splitting $E_{\text{VS}}$.
\textbf{e}, Projection of the unstrained cubic unit cell onto the $x$-$y$ plane. \textbf{f}, Schematic deformation of the unit cell under shear strain along [110]. \textbf{g}, Silicon tetrahedral bonds in the absence of strain. 
\textbf{h}, Shear strain affects bonds differently, depending on the crystal plane, causing a lattice deformation along [001]. \textbf{i}, Shear strain reduces the translation symmetry of the 1D lattice, resulting in a primitive unit cell of length $a_0/2$. \textbf{j}, The corresponding BZ is then folded in half, and the shear-strain coupling $k_z=4\pi/a_0$ is manifested as avoided crossings at the BZ boundary. Here, the green and blue arrows correspond to the same arrows in \textbf{c}, and the two black dots denote equivalent $k_z$ separated by a reciprocal lattice vector $4\pi/a_0$. Notice that the shear strain coupling (green arrow) conserves momentum in the reduced BZ.
(\textbf{c} and \textbf{j} are both calculated using the sp$^3$d$^5$s$^*$ tight-binding model described in Methods.)
\textbf{k}, Typical Ge concentration profile of a Wiggle Well heterostructure~\cite{McJunkin2021,Feng2022}. 
Here, the Ge oscillation wavelength, $ \lambda = \pi/k_1 \approx 1.68~\text{nm}$, is chosen to couple the valleys, as in \textbf{c}. 
\textbf{l}, \textbf{m}, Cartoon depiction of shear strain ($\varepsilon_{xy}$) in a quantum well, before (\textbf{l}) and after (\textbf{m}) mechanical bending. 
\textbf{n}, \textbf{o}, Cartoon depiction of shear strain induced by lithographic etching, before (\textbf{n}) and after (\textbf{o}) lattice deformation caused by cooling to cryogenic temperatures.}
\label{FIG1}
\vspace{-1mm}
\end{figure*}
%%%%%%%%%%%%%%%%%%%%%%	
%%%%%%%%%%%%%%%%%%%%%%%%%%%%%%%%

\section{Results}
\subsection{Valley splitting enhancement mechanism}
Shear strain has a subtle but profound effect on the silicon crystal lattice shown in Fig. \ref{FIG1}\textbf{a}.
In Fig. \ref{FIG1}\textbf{b}, we also show an effective 1D lattice obtained by assuming translational symmetry along $\hat x$ and $\hat y$ and performing a Fourier transformation.
(This 1D model forms the starting point for the effective-mass theory described below.)
The conventional primitive unit cell of Si contains two atoms (red and blue), which give rise to the two sublattices shown in the figure.
Now, for the special case of $k_x\!=\!k_y\!=\!0$, silicon possesses a screw symmetry, which interchanges the red and blue lattices, and reduces the primitive cell to one atom~\cite{Woods2023a}. 
Note that the weak in-plane confinement of a quantum dot has an insignificant effect on the valley coupling described here, justifying the focus on $k_x\!=\!k_y\!=\!0$.
The Brillouin zone along $\hat k_z$ can then be expanded from $k_z=\pm 2\pi/a_0$ to $\pm 4\pi/a_0$, as shown in Fig. \ref{FIG1}\textbf{c}.
Here $a_0$ is the size of the cubic unit cell, and the low-energy valley minima are located at $k_z=\pm k_0$.
In the absence of valley coupling, each state from the $-k_0$ valley will have a degenerate partner from the $k_0$ valley, as shown in the left-hand side of Fig. \ref{FIG1}\textbf{d}. Hence, a coupling between these valleys is needed for a valley splitting $E_{\text{VS}}$ between the two lowest energy states to exist.
A short-period WW -- if it could be grown -- would provide a direct (i.e., first-order) $2k_0$ coupling between the valleys, by introducing ultrashort-period Ge concentration oscillations into the quantum well (Fig. \ref{FIG1}\textbf{k}), with $\lambda=2\pi/(2k_0)$~\cite{McJunkin2021,Feng2022}.
Here, we propose a more physically realistic, second-order coupling scheme involving (i) shear strain, which couples $k_z$ states separated by $4\pi/a_0$ (green arrow in Fig. \ref{FIG1}\textbf{c}), and (ii) a long-period WW of wavelength $\lambda=2\pi/(2k_1)$ (blue arrow in Fig. \ref{FIG1}\textbf{c}).

The mechanism responsible for the strain-induced $4\pi/a_0$ coupling is illustrated in Figs.~\ref{FIG1}\textbf{e}-1\textbf{h}.
Here, shear strain $\varepsilon_{xy}$ is seen to elongate and compress the crystal along the $[110]$ and $[1\bar{1}0]$ directions, respectively.
In turn, this distorts the tetrahedral bonds of the diamond lattice and shifts the blue sublattice downward with respect to the red sublattice (along $[00\bar 1]$), as shown in Figs.~\ref{FIG1}\textbf{h} and \ref{FIG1}\textbf{i}.
This layer pairing causes the primitive cell to expand from one to two atoms, and the BZ folds back to the conventional boundaries at $k_z=\pm 2\pi/a_0$ (Fig.~\ref{FIG1}\textbf{j}).
The period of the distortion is $a_0/2$, which gives the desired $4\pi/a_0$ coupling vector upon Fourier transformation, and produces an avoided crossing of the low-energy bands, as observed at the zone boundary.
We note that the coupling responsible for this avoided crossing is the same as the green arrow in Fig.~\ref{FIG1}\textbf{j}, except it occurs between states at $k_z = \pm2\pi/a_{0}$ instead of $\pm k_{0}$.

The second-order process illustrated in Fig. \ref{FIG1}\textbf{c} couples the two valleys through a virtual (i.e., high-energy) state, indicated by a black dot.
The resulting momentum loop can be expressed as $2 k_1 = 4 \pi/a_0 - 2 k_0$, where the Ge concentration period $\lambda=\pi/k_1\approx 1.68~\text{nm}$ corresponds precisely to the long-wavelength WW.
We note that this period is 5.3 times larger than the period for the short-wavelength WW~\cite{Feng2022}, making it experimentally feasible~\cite{McJunkin2021}. 
We also note that such large strains are commonly employed in the microelectronics industry~\cite{Auth2008,Packan2008}, where they are used to improve transistor performance.
In the laboratory, we envision implementing shear strain through simple mechanical deformations, like those illustrated in Figs.~\ref{FIG1}\textbf{l} and 1\textbf{m}, for near-term experiments.
In the long term, scalable solutions will likely involve etched geometries, like those illustrated in Figs.~\ref{FIG1}\textbf{n} and 1\textbf{o}, or other stressor-based strategies used in industry. 
Below, we simulate several shear-strain geometries that could be implemented in experiments, obtaining strain levels sufficient for achieving deterministically enhanced valley splittings greater than 100~$\mu$eV.
These results indicate that the STRAWW proposal is feasible and can be achieved with existing technology.

%%%%%%%%%%%%%%%%%%%%%%%%%%%%%%%%
%%%%%%%%%%%%%%%%%%%%%%%%%%%%
\begin{figure}[t]
\begin{center}
\includegraphics[width=0.48\textwidth]{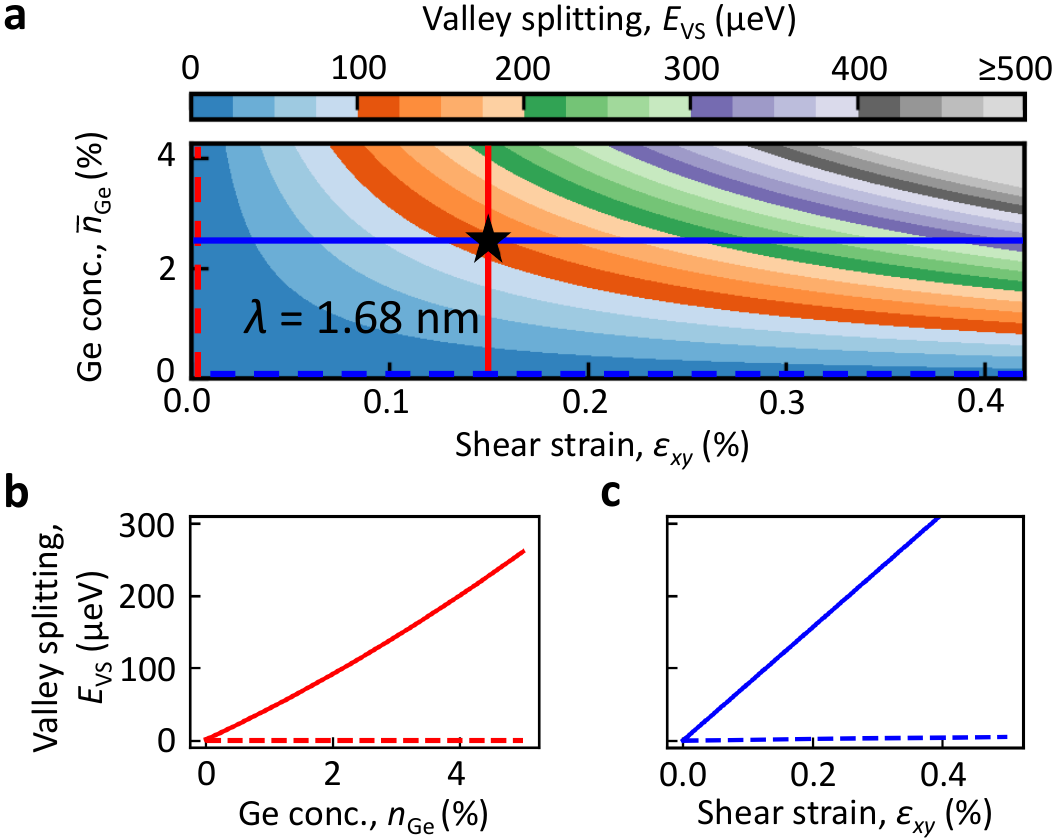}
\end{center}
\vspace{-0.5cm}
\caption{\textbf{Enhanced valley splitting.}
\textbf{a}, Valley splitting as function of shear strain and Ge concentration at the optimal WW wavelength $\lambda = 1.68~\text{nm}$, computed using sp$^3$d$^5$s$^*$ tight-binding theory. 
Here we assume a wide interface, $w = 1.9~\text{nm}$, and electric field, $F_z = 2~\text{mV/nm}$. 
(Black star refers to Fig.~\ref{FIG2_2}.) 
\textbf{b}, Line cuts from \textbf{a}, corresponding to $\varepsilon_{xy} = 0$ (dashed line) and $0.15\%$ (solid line). 
\textbf{c}, Line cuts from \textbf{a}, corresponding to $\bar{n}_{\text{Ge}} = 0$ (dashed line) and $2.5\%$ (solid line).
Only the \textit{combination} of shear strain and WW is found to significantly enhance
the valley splitting.}
\label{FIG2}
\vspace{-1mm}
\end{figure}
%%%%%%%%%%%%%%%%%%%%%%	
%%%%%%%%%%%%%%%%%%%%%%%%%%%%%%%%

\subsection{sp\textsuperscript{3}d\textsuperscript{5}s\textsuperscript{*} tight-binding calculations}
We now perform numerical simulations to quantify the effects of shear strain and Ge oscillations on the valley splitting, finding that significant enhancements are observed only when both features are present.
We begin by employing an sp$^3$d$^5$s$^*$ tight-binding model \cite{Niquet2009,Jancu1998}, which is known to give accurate results for the band structure over a wide range of energies (see Methods). 
The model incorporates tight-binding parameters for Si-Si, Ge-Ge, and Si-Ge nearest-neighbor hoppings, and can therefore describe arbitrary $\text{Si}_{1-x}\text{Ge}_{x}$ alloys.
The model also includes strain, yielding results that are in good agreement with experimentally measured deformation potentials~\cite{Niquet2009}.
Here, we focus on the deterministic component of $E_{\text{VS}}$, ignoring local fluctuations of the Ge concentration~\cite{Losert2023}.
The $\text{Si}\text{Ge}$ alloy is therefore treated in a virtual crystal approximation, following Ref.~\cite{Woods2023a}, where the Hamiltonian matrix elements are averaged over all possible alloy realizations.

In Fig.~\ref{FIG2}, we plot the valley splitting $E_{\text{VS}}$ as a function of the shear strain $\varepsilon_{xy}$ and the average Ge concentration in the quantum well $\bar{n}_{\text{Ge}}$.
We consider a uniform vertical electric field of strength $F_z = 2$~mV/nm and the optimal STRAWW oscillation wavelength of $\lambda = 1.68~\text{nm}$. 
We also assume a wide interface of width $w =1.9~\text{nm}$, as defined in Methods, 
which strongly suppresses the interface-induced valley splitting~\cite{Losert2023} and ensures that the valley splitting enhancements we observe here are caused by the STRAWW structure. 
Figure~\ref{FIG2}\textbf{a} shows results in the two-dimensional parameter space, while Fig.~\ref{FIG2}\textbf{b} shows two vertical linecuts, and Fig.~\ref{FIG2}\textbf{c} shows two horizontal linecuts, including the cases with $\varepsilon_{xy}=0$ or $\bar{n}_{\text{Ge}}=0$.
These results clearly indicate that a \textit{combination} of shear strain and concentration oscillations is needed to produce a large valley splitting.
Indeed, when $\varepsilon_{xy}=0$ or $\bar{n}_{\text{Ge}}=0$, we observe dangerously low values of $E_{\text{VS}} < 25~\mu\text{eV}$, over the whole parameter range. 
In contrast, when both $\varepsilon_{xy},\bar{n}_{\text{Ge}}>0$, we quickly approach a range of acceptable valley splittings. 
For example, when $\varepsilon_{xy}=0.15\%$ and $\bar{n}_{\text{Ge}}=2.5\%$ (black star in Fig.~\ref{FIG2}\textbf{a}), we obtain $E_{\text{VS}} \approx 125~\mu\text{eV}$.

%%%%%%%%%%%%%%%%%%%%%%%%%%%%%%%%
%%%%%%%%%%%%%%%%%%%%%%%%%%%%
\begin{figure}[t]
\begin{center}
\includegraphics[width=0.48\textwidth]{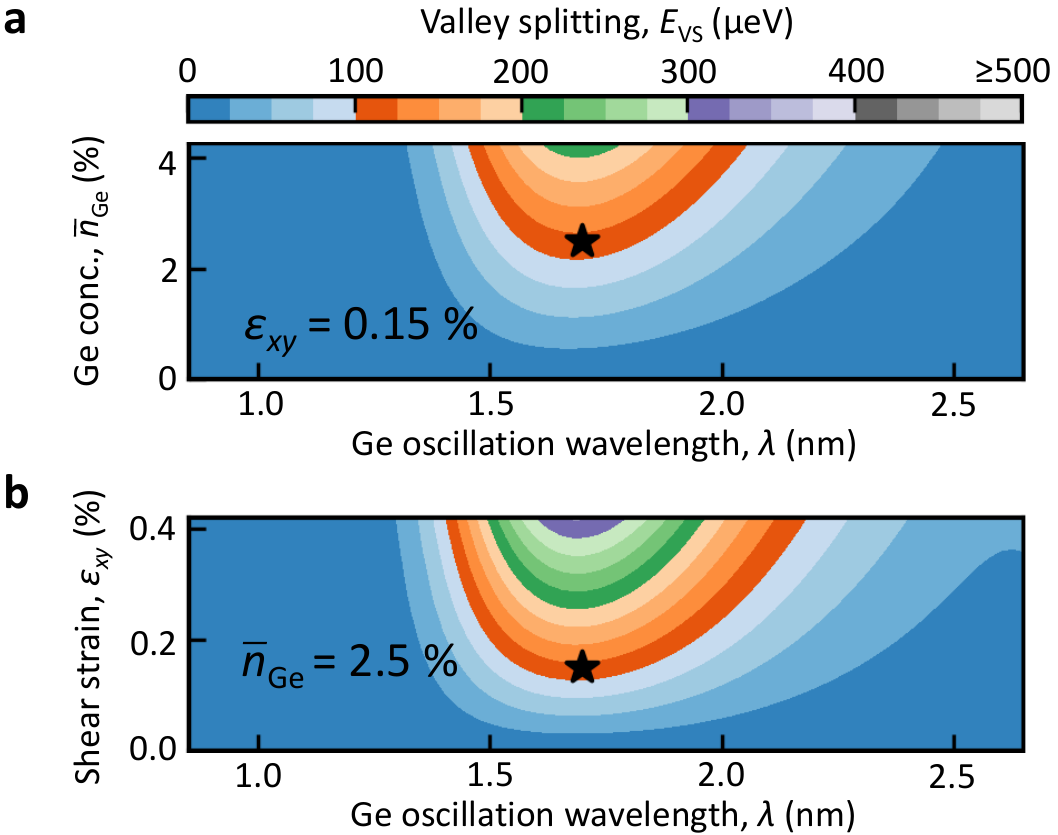}
\end{center}
\vspace{-0.5cm}
\caption{\textbf{Robustness of STRAWW to $\lambda$ errors.} 
Valley splitting as a function of the Ge oscillation wavelength $\lambda$, for fixed values of \textbf{a}, $\varepsilon_{xy} = 0.15\%$, or \textbf{b}, $\bar{n}_{\text{Ge}} = 2.5\%$.
For realistic device parameters (e.g., the black stars, referring to Fig.~\ref{FIG2}), acceptable valley splittings are achieved over a wide range of $\lambda$, providing robustness against growth imperfection.}
\label{FIG2_2}
\vspace{-1mm}
\end{figure}
%%%%%%%%%%%%%%%%%%%%%%	
%%%%%%%%%%%%%%%%%%%%%%%%%%%%%%%%

In Fig.~\ref{FIG2_2}, we show that the valley splitting enhancements observed in Fig.~\ref{FIG2} are robust to imperfect implementation of the STRAWW wavelength $\lambda=1.68$~nm.
Plotting $E_\text{VS}$ as a function of $\lambda$ and $\bar{n}_\text{Ge}$ in Fig.~\ref{FIG2_2}\textbf{a}, or $\varepsilon_{xy}$ in Fig.~\ref{FIG2_2}\textbf{b}, we find that a broad range of $\lambda$ values gives acceptable valley splittings. 
For example, at the physically realistic device setting indicated by a black star (the same setting as Fig.~\ref{FIG2}), we obtain $E_\text{VS}\approx 125$~$\mu$eV; on either side of this point, the range of wavelengths with $E_\text{VS}>100$~$\mu$eV extends from $\lambda=1.5$-1.9~nm, which is well within the growth tolerance achieved in Ref.~\cite{McJunkin2021}.
Finally, we note that, near the optimal wavelength $\lambda=1.68$~nm,  $E_{\text{VS}}$ varies linearly with both $\bar n_\text{Ge}$ and $\varepsilon_{xy}$, which we now explain using effective-mass theory. 

%%%%%%%%%%% Effective Mass Section
\subsection{Effective mass theory}
To gain further insight into the separate (or combined) contributions to the valley splitting from shear strain and Ge concentration oscillations, we outline here an effective-mass theory, with details given in Supplementary Note~1 \cite{SM}.
This formalism provides intuition regarding valley-splitting enhancement mechanisms arising from Ge concentration oscillations and shear strain, while also being amenable to analytic analysis.
Treating the valleys centered at $k_z = \pm k_0$ as pseudospins, the Hamiltonian takes the form
\begin{equation}
    H_{\text{EM}} = 
    \begin{pmatrix}
        H_{0} & H_v \\
        H_v^\dagger & H_{0}
    \end{pmatrix}, \label{HEM}
\end{equation}
where $H_{0}$ is the intra-valley Hamiltonian and $H_v$ describes the inter-valley couplings. 
The intra-valley Hamiltonian is given by
\begin{equation}
    H_{0} = -\frac{\hbar^2}{2 m_l} \partial_z^2 + V(z), \label{H0}
\end{equation}
where $m_l=0.92 m_e$ is the longitudinal effective mass, and 
\begin{equation}
    V(z) = V_0 \cos(\frac{2\pi z}{\lambda}) + V_{\text{qw}}(z),
    \label{eq:V0}
\end{equation}
where the first term describes potential modulations of amplitude $V_0$ due to the Ge oscillations, and $V_{\text{qw}}$ describes both the quantum-well confinement potential and the applied electric field. 
The inter-valley Hamiltonian takes the form 
\begin{equation}
\begin{split}
    H_v =& V(z) e^{-i 2 k_0 z}  \\
    &- i \varepsilon_{xy} e^{i 2 k_1 z}\left[
    C_1  
    + C_2  (i 2 k_1 \, \partial_z+\partial_z^2 )
    \right], 
\end{split}  \label{HinterValley}    
\end{equation}
where $C_1 = 1.73~\text{eV}$ and $C_2=0.067 a_0^2 C_1$ are constants, and $C_1$ is adjusted to match the valley splitting results from our sp$^3$d$^5$s$^*$ calculations. 
It is interesting to see two types of ``fast'' oscillations in this equation (indicated in Fig. \ref{FIG1}\textbf{c}), with wavevectors $2k_0$ arising from the potential, and $2k_1$ arising from the shear strain. 

We solve for the low-energy states of Eq.~(\ref{HEM}) using an envelope function method similar to \cite{Burt1988} (see also Methods) by treating $V_0$ and $C_1\varepsilon_{xy}$ perturbatively. Defining $\psi$ as the ground state of $H_0$ in Eq.~(\ref{H0}), this analysis yields a simple but extremely useful expression for the inter-valley matrix element, $\Delta= \mel{\psi}{H_v}{\psi}$:
\begin{equation}
\begin{split}
    \Delta = 
    \int |&F_{0}(z)|^2 \Big[ V_\text{qw}(z) e^{-i 2 k_0 z} 
     +\frac{V_0}{2} e^{-i (2 k_0 - G_\lambda)z} \\
    & -i \varepsilon_{xy} C_1 e^{i 2 k_1 z} \\
   &-i \varepsilon_{xy}\frac{V_0}{E_\lambda}
   \left(C_1 - C_2 k_1 G_\lambda\right)
   e^{i (2 k_1 - G_{\lambda}) z}
    \Big] \,dz ,
\end{split} \label{DeltaMtxElem}
\end{equation}
where $F_0(z)$ is the envelope function of $\psi(z)$ (obtained by setting $V_0=0$),
and we have defined the oscillation wavevector and oscillation energy scales as $G_{\lambda} = 2\pi/\lambda$ and $E_\lambda = 2 \hbar^2 \pi^2/(m_l \lambda^2)$, respectively.
Note here that $E_{\text{VS}} = 2|\Delta|$.

Each of the four terms in Eq.~(\ref{DeltaMtxElem}) effectively averages to zero for smoothly varying $F_0$ and $V_\text{qw}$, except when specific resonance conditions are met, and each term has a distinct physical origin and meaning, as explained below. 
The first term is the usual inter-valley matrix element in the absence of Ge oscillations and shear strain~\cite{Losert2023}.
The second term describes Ge oscillations, independent of shear strain, and is the basis for the short-wavelength WW~\cite{McJunkin2021,Feng2022}. 
The third term describes the shear strain contribution, independent of Ge oscillations.
The fourth term involves both shear strain and Ge oscillations, and is unique to STRAWW. 
The resonance condition for which this term has a vanishing exponential,
$2 k_1 \approx G_\lambda$, corresponds to the long-period WW, or $\lambda = \pi/k_1 \approx 1.68~\text{nm}$.

%%%%%%%%%%%%%%%%%%%%%%%%%%%%%%%%
%%%%%%%%%%%%%%%%%%%%%%%%%%%%
\begin{figure*}[t]
\begin{center}
\includegraphics[width=.98\textwidth]{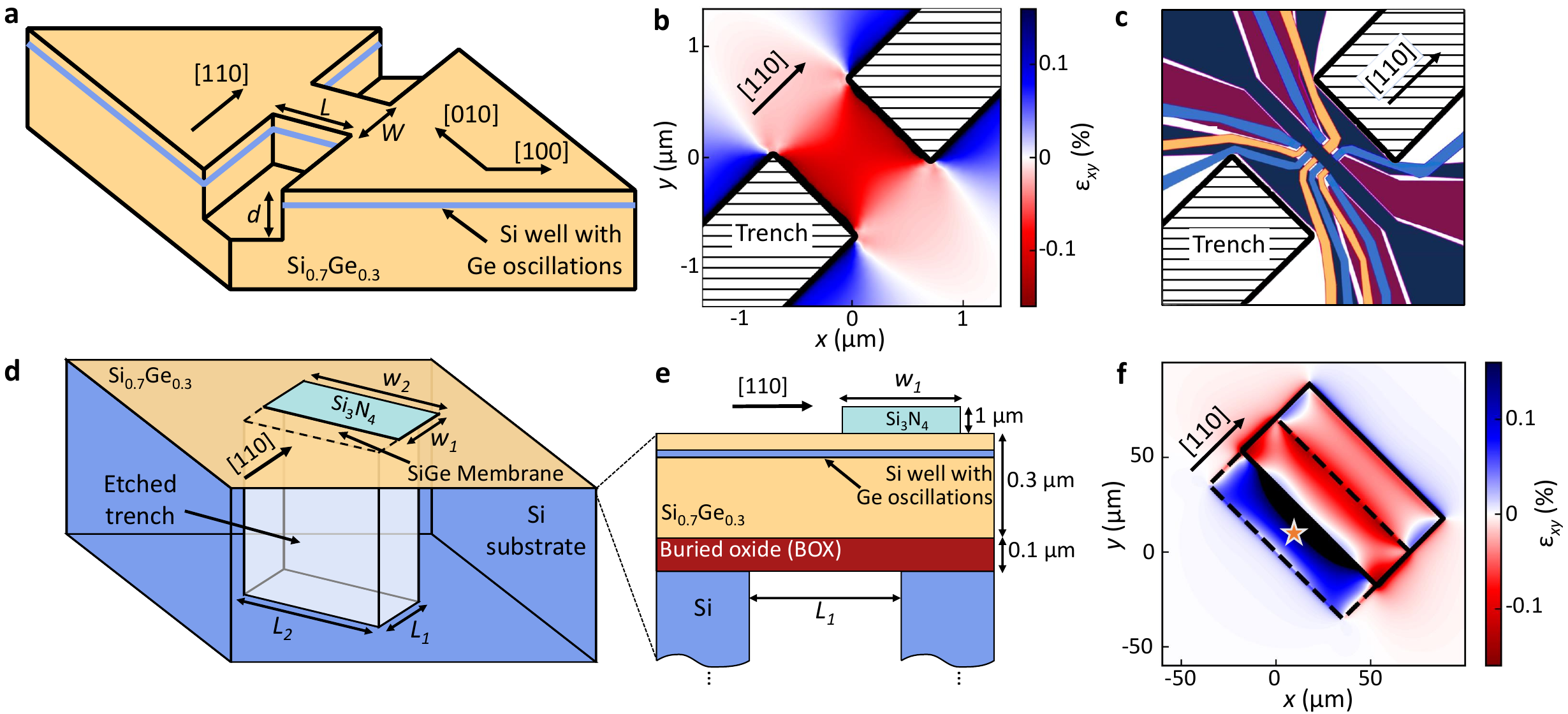}
\end{center}
\vspace{-0.5cm}
\caption{\textbf{Strategies for implementing shear strain.} 
\textbf{a}, Microbridge geometry of dimension $L \cross W$, defined by two trenches of depth $d$ etched into the top of a Si/SiGe heterostructure along the $[110]$ direction. 
\textbf{b}, Calculated shear strain $\varepsilon_{xy}$ in the quantum well, for the geometry defined by $L = W = d = 1~\mu\text{m}$. 
A large shear strain, $\varepsilon_{xy} \approx -0.1\%$, is obtained in the center of the channel.
\textbf{c}, A triple-dot gate design, analogous to Ref.~\cite{Dodson2022}, with gates arranged to avoid the trenches, for the same geometry as \textbf{b}. 
\textbf{d}, A 3D membrane-stressor geometry, with \textbf{e}, a corresponding [110] cross-section. 
A free-standing membrane is formed by etching a trench, with lateral dimensions $L_1\times L_2$, from the bottom of a Si substrate to a buried oxide (BOX) etch stop.
A Si$_3$N$_4$ stressor of dimension $w_1\times w_2$ is fabricated atop the Si/SiGe quantum well.
The stressor is centered above one trench edge, such that it partially covers the membrane.
Both membrane and stressor are aligned along [110].
\textbf{f}, Calculated shear strain in the quantum well, for the geometry defined by $(L_{1}, L_{2}, w_{1}, w_{2})=( 50, 100, 50, 100)~\mu\text{m}$, where the membrane and stressor regions are outlined by dashed and solid lines, respectively. The membrane deforms oppositely, depending on whether it is covered or uncovered by the stressor.
Relatively uniform strain is achieved across the wide, blue region, with $\varepsilon_{xy} = 0.15\%$ at the location of the orange star.}
\label{FIG4}
\vspace{-1mm}
\end{figure*}
%%%%%%%%%%%%%%%%%%%%%%	
%%%%%%%%%%%%%%%%%%%%%%%%%%%%%%%%

We can evaluate Eq.~(\ref{DeltaMtxElem}) for a specific quantum well model.
For simplicity, we treat the quantum well as a harmonic confining potential with $V_{\text{qw}} = m_l \omega_z^2 z^2/2$, and the envelope function $F_0(z) = (\pi l_z^2)^{-1/4}\exp(z^2/(2l_z^2))$, where $l_z = \sqrt{\hbar/(m_l \omega_z)}$ is the characteristic dot size in the growth direction. 
Equation~(\ref{DeltaMtxElem}) then reduces to
\begin{equation}
\begin{split}
    \Delta =& 
    \frac{\hbar^2 k_0^2}{2 m_l} e^{-(k_0 l_z)^2}
    + \frac{V_0}{2} e^{-(2k_0-G_\lambda)^2 l_z^2/4} \\
    &- i \varepsilon_{xy} C_1 e^{-(k_1 l_z)^2}\\
    &- i \varepsilon_{xy}(C_1 - C_2 k_1 G_\lambda) \frac{V_0}{E_\lambda} e^{-(2k_1-G_\lambda)^2 l_z^2/4}.
\end{split} \label{DeltaHarmonic}
\end{equation}
If we now assume a long-period WW, with $2k_1=G_\lambda$, the first three terms in Eq.~(\ref{DeltaHarmonic}) are all exponentially suppressed, leaving $E_\text{VS}\approx (2\varepsilon_{xy}V_0/E_\lambda)(C_1 - 2 C_2 k_1^2)$.
This confirms the anticipated linear dependence on $\varepsilon_{xy}$ and $V_0$, and provides a simple estimate for the STRAWW valley splitting.
We also note that this expression is independent of $l_z$, justifying our approximate treatment of the quantum well confining potential, and indicating that the valley splitting does not depend on the interface for this smooth quantum-well geometry.

In the opposite limit of an ultra-sharp interface, Eq.~(\ref{DeltaHarmonic}) is no longer accurate.
In this case, the singular nature of the confining potential $V_\text{qw}$ (or the envelope function $F_0$) generates Fourier components that cancel out the fast-oscillating phase factors in Eq.~(\ref{DeltaMtxElem}), so the first three terms are not suppressed.
The resulting valley splitting enhancement has been studied previously for quantum wells with~\cite{Ungersboeck2007,Sverdlov2008,Adelsberger2023} and without~\cite{Losert2023} shear strain.
However in the latter case, the crossover between ultra-sharp vs smooth interfaces occurs for interface widths of just 1-2 atomic monolayers, which are extremely difficult to grow in the laboratory~\cite{Wuetz2021}, so valley splitting enhancements are rarely observed.
In Supplementary Note~2 \cite{SM}, we show numerically that the same is true for shear-strained structures, suggesting that the STRAWW strategy is much more practical.

\subsection{Strain calculations}
Finally, we show that the shear strains needed to deterministically enhance the valley splitting can be achieved with current micro and nanofabrication technologies.
In Fig.~\ref{FIG4}, we consider two etching strategies. 
Figure~\ref{FIG4}\textbf{a} shows a microbridge geometry~\cite{Suess2013,Minamisawa2012}, formed by etching the top surface to well below the quantum well.
Here, the trenches are aligned along the $[110]$ crystallographic direction to provide shear strain in the channel region between the trenches. 
We calculate the strain in the quantum well numerically, using COMSOL Multiphysics~\cite{Comsol}, for the geometry shown in Fig.~\ref{FIG4}\textbf{a}, with $L = W = d = 1~\mu\text{m}$.
The resulting shear strain $\varepsilon_{xy}$ shown in Fig.~\ref{FIG4}\textbf{b} is largely uniform across the channel, except for edge strips with an approximate width of 100~nm.
At the center of the channel, we obtain $\varepsilon_{xy} \approx -0.1\%$, which yields a large $E_{\text{VS}}$, even for a moderate WW amplitude (see Fig.~\ref{FIG2}). 
In this geometry, the trenches are sufficiently separated to allow for electrode fabrication using the realistic design shown in Fig.~\ref{FIG4}\textbf{c}. This design has geometric parameters comparable to Ref.~\cite{Dodson2022} in the dot region, and here we exploit the large amount of free space in the orthogonal direction to fan out the gates, to avoid forming sharp kinks that could lead to lithographic failure.
To allow for a higher density of gate electrodes, we envision filling the trenches with an oxide material, to restore a planar top surface on which additional gates can be fabricated.

Figures~\ref{FIG4}\textbf{d} and 4\textbf{e} show a second, stressor-type geometry, featuring a Si/SiGe heterostructure grown atop a buried oxide (BOX) layer~\cite{Mizuno2002}, which serves as an etch-stop for a trench etched from the bottom of the Si substrate. 
The latter is aligned along [110], creating a free-standing Si/SiGe membrane~\cite{Guo2018,Chavez2014}, which is readily strained by a $\text{Si}_{3}\text{N}_{4}$ stressor~\cite{Jain2012,Ghrib2012,Toivola2003} (also aligned along [110]), due to the thinness of the membrane. 
For the geometry defined by $(L_{1}, L_{2}, w_{1}, w_{2})= (50, 100, 50, 100)~\mu\text{m}$, we obtain the strain results shown in Fig.~\ref{FIG4}\textbf{f}, where we assume a depositional tensile stress of $1~\text{GPa}$ for the $\text{Si}_3 \text{N}_4$ stressor~\cite{Jain2012}. 
Here, the red region is covered by stressor, while the blue region is uncovered, but located above the trench.
This blue region, which is most convenient for fabricating gates, provides a high and largely uniform shear strain, with $\varepsilon_{xy} \approx 0.15\%$ at the location of the orange star, and $\varepsilon_{xy} \geq 0.15\%$ over a wide area ($>785~\mu \text{m}^2$) that could potentially contain many dots in a dense two-dimensional array~\cite{Borsoi2022,Hendrickx2021,Mortemousque2021,Unseld2023}.

\section{Discussion}
In summary, we have shown that the combination of shear strain and the long-period Wiggle Well, dubbed ``STRAWW,'' is particularly effective for deterministically enhancing the valley splitting, which is a key missing ingredient for scaling up silicon spin qubits to large arrays. 
Effective-mass theory shows that this enhancement arises from a second-order coupling process that requires breaking the screw symmetry of the diamond lattice.
Numerically accurate tight-binding simulations suggest that valley splittings needed for qubit operation can be achieved using realistic shear strains ($\varepsilon_{xy}\approx 0.1$\%), while the required Ge concentration oscillations ($\lambda=1.7$~nm) have already been demonstrated in the laboratory~\cite{ McJunkin2021}.
While ``intrinsic'' strain sources such as metallic gates, dislocations, and dielectric layers do not provide a sufficiently large shear strain for our purposes~\cite{Thorbeck2015,Park2016,Frink2023,CorleyWiciak2023,ONeill2021}, our strain simulations confirm that several different etching techniques could be used to achieve such strain.
Overall, the fabrication requirements for STRAWW are much less challenging than other schemes, including ultra-sharp interfaces and short-period Wiggle Wells, making this a very attractive approach for managing valley splitting in future qubit experiments.
Moreover, spin-orbit coupling is also enhanced in long-period Wiggle Wells, which may be exploited for qubit manipulation~\cite{Woods2023a}.
We expect that future devices could provide shear strains larger than those reported here, through a variety of implementation strategies.

%TC:ignore
\section{Methods}

\subsection{sp\textsuperscript{3}d\textsuperscript{5}s\textsuperscript{*} tight-binding model}

We use an sp$^3$d$^5$s$^*$ tight-binding model~\cite{Niquet2009} to study valley coupling in Si/SiGe quantum wells. 
This is an atomistic model with nearest-neighbor hopping terms, including ten spatial orbitals at each atom site. 
Such models are well established for accurately modeling the electronic structure of many different semiconductors~\cite{Jancu1998}. 
Spin-orbit coupling is typically introduced into these tight-binding models through an intra-atomic coupling between p orbitals \cite{Chadi1977}.
However, this has no quantitative effect on the valley splitting results studied here, and is therefore disabled in the current analysis, for simplicity.
On the other hand, we incorporate distinct Si-Si, Ge-Ge, and Si-Ge nearest-neighbor hopping terms, following~\cite{Niquet2009}, to more accurately represent SiGe alloys. 

Strain is incorporated into the model through the equation
\begin{equation}
    \boldsymbol{R}_i = \left(1 + \boldsymbol{\varepsilon}\right)  \boldsymbol{R}_{i,0}
    \pm \zeta \frac{a_0}{4}\left(\varepsilon_{yz}, \varepsilon_{xz}, \varepsilon_{xy}\right), \label{AtomPos}
\end{equation}
which relates the location of atom $i$ in the presence of strain ($\boldsymbol{R}_i$) to its unperturbed location ($\boldsymbol{R}_{i,0}$).
Here, $\boldsymbol{\varepsilon}$ is the strain tensor, the signs $+$ and $-$ are assigned depending on whether atom $i$ belongs to the red or blue sublattice, respectively (see Fig. \ref{FIG1}\textbf{a}), and $\zeta$ is Kleinman's internal strain parameter~\cite{Kleinman1962}, which is used in diamond crystal lattices to account for the relative shift of the two sublattices within a unit cell due to shear strain.
We ignore the small difference between $\zeta$ values for Si vs Ge atoms, simply adopting $\zeta = 0.53$ throughout the structure~\cite{Guler2016}. 
Strain is then incorporated into the tight-binding Hamiltonian in two ways: (i) strain-dependent onsite couplings, following~\cite{Niquet2009},  and (ii) modified bond lengths and angles, which result in modified hopping terms~\cite{Slater1954,Froyen1979}.
Further details on our implementation of the sp$^3$d$^5$s$^*$ tight-binding model are provided in~\cite{Woods2023a}. 

We consider a sigmoidal-interface model \cite{Wuetz2021,Losert2023} with sinusoidal oscillations for the Ge concentration profile in the tight-binding calculations, given by 
\begin{equation}
    n_{\text{Ge}}(z) =  \frac{n_{\text{bar}}}{1 + e^{4z/w}}
    + 2\bar{n}_{\text{Ge}}\sin^2\left(\frac{\pi z}{\lambda}\right), \label{nGeProfile}
\end{equation}
where $n_{\text{bar}} = 0.3$ is the Ge concentration in the barrier region, $w$ is the interface width, $\bar{n}_{\text{Ge}}$ is the average Ge concentration in the quantum well (associated with the Ge concentration oscillations), and $\lambda$ is the oscillation wavelength. 
Note that the Ge concentration oscillations in Eq.~(\ref{nGeProfile}) extend across the heterostructure (including the barrier region). 
In realistic devices, the Ge oscillations would likely only occur in the quantum well region, as illustrated in Fig. \ref{FIG1}\textbf{k}. 
However, restricting the oscillations to the well region in the simulations produces kinks in $n_{\text{Ge}}$ profile, which are known to artificially boost the valley by enlarging the short-wavelength Fourier components of $n_{\text{Ge}}$~\cite{Losert2023}. 
Such kinks are not expected to be present in realistic devices.
Note that we have numerically verified that restricting the Ge concentration oscillations to the quantum well region does not significantly alter the valley coupling arising from the \textit{combination} of Ge concentration oscillations and shear strain.
We therefore avoid any such effects here by extending the Ge concentration oscillations across the whole structure. 
We also note that the bottom quantum well interface has been excluded in Eq.~(\ref{nGeProfile}), because the electric field used in our simulations is always large enough to push the electron wave function tightly against the top interface.
For narrower quantum wells, where this would not be the case, we do not expect the valley splitting provided by STRAWW to be significantly affected by including a bottom interface. This is because the valley coupling within STRAWW does not fundamentally rely upon interface effects, in contrast to ``conventional'' Si/SiGe quantum wells without Ge concentration oscillations.

\subsection{Envelope-function method} \label{EFM}

Here, we use an envelope-function approach to derive Eq.~(\ref{DeltaMtxElem}) of the main text within the effective-mass framework of Eq.~(\ref{HEM}). To begin, we note that the energy scale of the intra-valley Hamiltonian $H_0$ dominates over the inter-valley Hamiltonian $H_v$. 
We can therefore calculate $E_{\text{VS}}$ by first solving for the ground state of $H_0$, and then computing the matrix element of $H_v$ within first-order perturbation theory. 

We can solve $H_0$ straightforwardly with an effective-mass approximation.
First, we note that the ground state $\psi$ of $H_0$ (or any other eigenstate) may be expanded as~\cite{Burt1988}
\begin{equation}
    \psi(z) = \sum_{n} F_{n}(z) U_n(z),\label{EnvExp}
\end{equation}
where $\{U_n\}$ comprises a complete and orthogonal set of Bloch functions that satisfy the $k = 0$ Bloch-Schr\"{o}dinger equation,
\begin{equation}
    \left[
    -\frac{\hbar^2}{2 m_l} \partial_z^2 + V_0
    \cos\left(\frac{2\pi z}{\lambda}\right)\right] U_n(z) = E_n U_n(z),
\end{equation}
and $F_n$ are envelope functions that account for the effects of the confinement, $V_{\text{qw}}$. 
Note that the Bloch functions here do not have the periodic structure of the crystal lattice, but rather the periodic structure of the WW, which enters $H_0$ through the potential $V(z)$ in Eq.~(\ref{eq:V0}).
$U_n$ are therefore periodic over the length scale $\lambda$, while $F_n$ are slowly varying over this length scale. 
Here we adopt the conventional normalizations $\lambda^{-1}\int_0^\lambda |U_n(z)|^2dz=1$ and $\sum_n\int|F_n(z)|^2dz=1$.
Plugging Eq.~(\ref{EnvExp}) into the Schr\"{o}dinger equation yields the coupled envelope equations
\begin{equation}
    \begin{split}
        -\frac{\hbar^2}{2m_l}&\partial_z^2 F_n 
        - \sum_m \frac{i \hbar}{m_l} p_{nm} \partial_z F_m(z) \\
        &+ \sum_m \int V_{nm}(z,z^\prime) F_m(z^\prime) \,dz^\prime 
        = (E - E_n) F_n(z),
    \end{split} \label{EnvEq}
\end{equation}
where 
\begin{equation}
    p_{nm} = \frac{-i\hbar}{\lambda}\int_0^\lambda U_n^* \partial_z U_m \,dz, 
\end{equation}    
and $p_{nn}=0$, and $V_{nm}(z,z^\prime)$ is a non-local potential arising from $V_{\text{qw}}$, whose form is given in Ref.~\cite{Burt1988}. 

Up to this point, the expansion of Eq.~(\ref{EnvExp}) and the following equations are exact.
However, the Schr\"odinger-like, coupled envelope equations, Eq.~(\ref{EnvEq}), can be effectively decoupled via a canonical transformation, yielding the much simpler result, $\psi(z)\approx F_0(z)U_0(z)$, where $F_0(z)$ satisfies an effective mass equation.
To derive this result, we note that the non-local and inter-band nature of the potential $V_{nm}(z,z')$ arises from the Fourier components of $V_{\text{qw}}$ in the outer-half of the Brillouin zone, i.e.\ for large wavevectors $|k| > \pi/(2\lambda)$. 
Assuming that $V_{\text{qw}}$ varies slowly compared to $\lambda$, these Fourier components are insignificant, and we can ignore all non-local and inter-band components in $V_{nm}(z,z^\prime)$.
In other words, there is a separation of length scales leading to the following \textit{local} approximation \cite{Burt1988} for Eq.~(\ref{EnvEq}):
\begin{equation}
    \begin{split}
        -\frac{\hbar^2}{2m_l}\partial_z^2 F_n 
        &-\frac{i \hbar}{m_l} \sum_m p_{nm} \partial_z F_m \\
        &+ V_{\text{qw}}(z) F_n(z) 
        = (E-E_n) F_n(z).
    \end{split} \label{EnvEq2}
\end{equation}

Now, there is also a large separation of energy scales that we can use to simplify the envelope equations.
The largest energy scale in Eq.~(\ref{EnvEq2}), by far, is the splitting between the different Bloch modes, $E_n - E_0\sim E_\lambda= 2 \hbar^2 \pi^2/(m_l \lambda^2)~(n\geq 1)$.
In the absence of coupling between these Bloch modes (i.e., for $p_{nm}=0$), the confinement energy scale arising from $V_\text{qw}$ is of order of $2 \hbar^2 \pi^2/(m_l l^2)\ll E_\lambda$, where $l$ is the width of the quantum well, while the inter-mode coupling term, which scales as $1/(l\lambda)$ is intermediate between these two energy scales.
A perturbative treatemenmt of the coupling is therefore justified.
Applying a Schrieffer-Wolff transformation, we obtain the desired single-mode solution $\psi(z)\approx F_0(z)U_0(z)$, where $F_0$ satisfies the effective-mass equation,
\begin{equation}
    -\frac{\hbar^2}{2m_l^*}\partial_z^2 F_0 + V_{\text{qw}}(z) F_0(z) \approx 
    (E - E_0)F_0(z) \label{FoEff} ,
\end{equation}
and the renormalized effective mass is given by
\begin{equation}
    \frac{1}{m_l^*} = \frac{1}{m_l} - 2 \frac{1}{m_l^2}\sum_{n \geq 1}
    \frac{\left|p_{n0}\right|^2}{E_n - E_0}.
\end{equation}

Equation~(\ref{FoEff}) is a standard effective-mass equation, where the periodic potential associated with the WW has been absorbed into the renormalized effective mass. 
However, the effective-mass correction (${\cal O}[V_0^2]$) is very small for typical heterostructures, since $V_0 \lesssim 20~\text{meV}$, while $E_\lambda \approx 565~\text{meV}$ for $\lambda = 1.7~\text{nm}$, such that $(m_l^*-m_l) / m_l  \lesssim 10^{-3}$. 
It is therefore a very good approximation to ignore the effective-mass renormalization. 
The leading-order correction in this formalism (${\cal O}[V_0]$) appears only in the Bloch function $U_0$, which to leading-order in $V_0$ is given by
\begin{equation}
    U_0 \approx 1 + (V_0/E_\lambda) \cos(2\pi z/\lambda). \label{U0}
\end{equation}

Finally, we are in a position to compute the inter-valley matrix element $\Delta= \mel{\psi}{H_v}{\psi}$ using $\psi(z) = F_0(z) U_0(z)$.
The result contains many terms, but is greatly simplified by ignoring the higher-order $\mathcal{O}[V_0^2]$ terms, terms involving derivatives of $F_0$, and integrals containing highly oscillatory plane waves that average toward zero. 
[Note that terms involving derivatives of the envelope function $F_0$ are insignificant, because the $F_n$ are slowly varying compared to the fast-oscillating $U_0$ terms.] 
A straightforward calculation then leads to Eq.~(\ref{DeltaMtxElem}) in the main text, and the corresponding valley splitting is given by $E_{\text{VS}} = 2 \left|\Delta\right|$. 

\subsection{Strain calculations}
Shear strain calculations were performed using the Solid Mechanics module in COMSOL Multiphysics~\cite{Comsol}.  
Below, we describe the materials properties assumed in these simulations, including the coefficients of thermal expansion (CTE), which describe thermal contractions when  the device is cooled from room temperature (293.15~K) to 1~K.

For the top-trench geometry of Fig.~\ref{FIG4}\textbf{a}, we assume a quantum well formed of a $9~\text{nm}$ strained-silicon layer sandwiched between a thick strain-relaxed buffer layer and a 50~nm top layer of strain-relaxed $\text{Si}_{0.7}\text{Ge}_{0.3}$. In a typical device, we would assume a buffer layer thickness of order 500~$\mu$m, which is challenging to simulate. 
In our simulations, we approximately account for such thick layers by assuming a thinner layer of width 2~$\mu$m (which is still much thicker than the top layers), and adopting a zero-displacement boundary condition, as defined in \cite{Comsol}.
The total volume of the resulting simulation is about
$3.2\times 10^{3}~\mu\text{m}^3$. 
Biaxial strain is applied to the Si layer through the initial-strain parameters $\varepsilon_{xx}=\varepsilon_{yy}=1.24\%$, and $\varepsilon_{zz}=-1.01\%$. 

The CTEs of Si and SiGe depend weakly on temperature, while COMSOL assumes constant CTE values.
To account for this in our simulations, we simply adjust the CTE values in the COMSOL parameter library.
Averaging over the CTEs provided in Refs.~\cite{Bradley2013,Goldberg2001,Corruccini1961} and applying Vegard's law to describe the SiGe alloy, we find the temperature-adjusted coefficients to be $\alpha_\text{Si}=0.76\times 10^{-6}~\text{K}^{-1}$ for Si, and $\alpha_{\text{Si}_{0.7}\text{Ge}_{0.3}}=1.52\times 10^{-6}~\text{K}^{-1}$ for the SiGe. 
For both Young's modulus $E$ and the Poisson ratio $\nu$, thermal corrections amount to only a few percent at most; we therefore use room temperature values for these parameters, as given in the COMSOL parameter library. 
To fully account for the $500~\mu\text{m}$ Si buffer layer (which is not included in the simulation), we also include the thermally corrected Si CTE, $\alpha_{\text{Si}}$, to describe the contraction of the bottom surface of the simulated device. 
To check the validity of our approximate treatment of the thick buffer layer, we also simulate thicker systems, to check for convergence, and we relax the boundary conditions by applying COMSOL's rigid-motion-suppression feature.

For the free-standing membrane geometry shown in Fig.~\ref{FIG4}\textbf{d}, we adopt the heterostructure parameters shown in Fig.~\ref{FIG4}\textbf{e}.
We also assume a trench height of $h = 500~\mu\text{m}$.
Here, we simulate a much larger total volume of about $5\times 10^{6}~\mu\text{m}^{3}$, to more accurately describe the effects of the deep trench. 
Biaxial strain is imposed as described above, and we account for depositional strain in the $\text{Si}_{3}\text{N}_{4}$ stressor by imposing an intrinsic tensile stress of $1~\text{GPa}$~\cite{Jain2012}. 
Here again we relax the boundary conditions using COMSOL's rigid-motion-suppression feature.
To account for thermal contraction, we take the same approach as above, with the additional temperature-adjusted CTEs given by $\alpha_{\text{Si}_{3}\text{N}_{4}}=0.5\times 10^{-6}\text{K}^{-1}$ and $\alpha_{\text{Si}\text{O}_{2}}=-0.2\times 10^{-6}\text{K}^{-1}$~\cite{Martyniuk2004,Middelmann2015,White1973}. 
We note that nearly identical shear-strain results can be obtained using room-temperature CTE values from the COMSOL library without accounting for cryogenic cooling, indicating that depositional strain from the $\text{Si}_{3}\text{N}_{4}$ stressor dominates over thermal contraction in this geometry.

\section*{DATA AVAILABILITY}
The data used to generate various figures is available at GitHub (\url{https://github.com/BenjaminWoods1/STRAWW}).

\section*{Code Availability}
The code used for the tight-binding and strain simulations is available at GitHub (\url{https://github.com/BenjaminWoods1/STRAWW}).

\section*{Acknowledgements}

We acknowledge helpful discussions with D.\ Savage. 
This research was sponsored in part by the Army Research Office (ARO) under Awards No.\ W911NF-17-1-0274, No.\ W911NF-22-1-0090, and No.\ W911NF-23-1-0110. 
The views, conclusions, and recommendations contained in this document are those of the authors and are not necessarily endorsed nor should they be interpreted as representing the official policies, either expressed or implied, of the ARO or the U.S. Government. The U.S. Government is authorized to reproduce and distribute reprints for Government purposes notwithstanding any copyright notation herein.

\section*{Author Contributions}
B.W. developed the idea of combining shear strain and Ge concentration oscillations and performed the tight-binding and effective mass calculations under the supervision of M.F. B.W, E.J., R.J., M.E, and M.F. developed the strategies for implementing shear strain. H.S. and C.F. performed the strain calculations. B.W., M.F., and H.S. wrote the manuscript with contributions from all authors.

\section*{Competing interests}
The Authors declare no Competing Non-Financial Interests but the following Competing Financial Interests. Authors B.W., E.J., R.J., M.E., and M.F. have applied for a patent on the STRAWW structure described here.

%\bibliography{References}

\clearpage

\renewcommand\thefigure{S.\arabic{figure}}
\renewcommand\theequation{S.\arabic{equation}}
\setcounter{figure}{0}
\setcounter{equation}{0}
\setcounter{section}{0}
\onecolumngrid
\vspace{0.5in}
\begin{center}
\textbf{\large Supplementary Information: Coupling conduction-band valleys in SiGe heterostructures via shear strain and Ge concentration oscillations}
\end{center}

\section{Supplementary Note 1}
In the Methods section of the main text, 
we showed how to analytically incorporate effects of the WW, using an effective-mass formalism.
In this Supplementary Note,
%Here, 
we show how to further incorporate the effects of shear strain into this theory, beginning with a minimal one-dimensional tight-binding model, where only a single orbital is included on each site. 

Ignoring effects of lateral confinement, a minimal, one-dimensional tight-binding model for a silicon quantum well can be written as  
\begin{equation}
\begin{split}
    \mel{m}{H_{\text{TB}}}{n} =& ~t_1 \delta_{m}^{n \pm 1} + t_2\delta_{m}^{n \pm 2} + \left(V_n - 2 t_1 - 2 t_2\right) \delta_{m}^{n} \\  
    %&+ \left(E_{\text{Ge}} n_{\text{Ge},n} + eF_z z_n \right) 
    &+(-1)^{n} \varepsilon_{xy} t_{\varepsilon}\left( 
    \delta_{m}^{n+1} - \delta_{m}^{n - 1}
    \right)
\end{split}, \label{HamTB}
\end{equation} 
where $\ket{n}$ is the orbital on site $n$ and $\delta$ is the Kronecker-delta function. The first line of Eq.~(\ref{HamTB}) is the same tight-binding Hamiltonian derived in Ref.~\cite{Boykin2004a}, where $t_1 = -0.60~\text{eV}$ and $t_2 = 0.62~\text{eV}$ are nearest and next-nearest neighbor hopping terms, respectively, and $V_n$ is an on-site, potential energy that captures the effects of external electric fields, as well as the quantum-well confinement potential associated with the Ge concentration profile.
Note that the magnitudes of $t_1$ and $t_2$ used here are slightly different than those reported in Ref.~\cite{Boykin2004a}, reflecting the slightly different locations of the valley minima $k_z=\pm k_0$ obtained in our sp$^3$d$^5$s$^*$ tight-binding model. 
In addition, we note the different sign used for $t_1$, as compared to Ref.~\cite{Boykin2004a}, which can be viewed as a gauge transformation in which the sign of the even-site orbitals is flipped. 
This gauge choice places the valley minima here at $k_z = \pm k_0$, whereas the valley minima sit at $k_z = \pm k_0 \pm 2k_1$ for the gauge choice of Ref.~\cite{Boykin2004a}.
Here, the Ge concentration profile is included within the virtual-crystal approximation. 
This model captures the essential features of the bottom of the Si conduction band, including the location of the valleys in momentum space at $k_z=\pm k_0 = \pm 0.83(2\pi/a_0)$, where $a_0$ is the size of the cubic unit cell, and the longitudinal effective mass is given by $m_l = 0.92 m_e$. 
(For simplicity, we henceforth take $k_z\rightarrow k$.)

The second line of Eq.~(\ref{HamTB}) is not present in Ref.~\cite{Boykin2004a}, and is included here to describe shear strain, $\varepsilon_{xy}$, where the new hopping parameter $t_\varepsilon$ is chosen as described below. 
Importantly, we note that the sign of the shear-strain hopping parameter alternates between neighboring sites, as a consequence of how the shear strain deforms the bonds of the Si crystal, as illustrated in Figs.~1(g) and 1(h) of the main text.
This alternating sign reflects the same underlying physics as the Kleinman strain term in Eq.~(7) of the main text, which we can understand as follows. 
A bulk Si crystal has two bond types, aligned along the $[111]$ and $[1\bar{1}1]$ directions, respectively, which alternate between atomic layers. 
Under shear strain, with $\varepsilon_{xy} > 0$, the system is elongated and shortened along the $[110]$ and $[1\bar{1}0]$ directions, respectively, as shown in Figs.~1(e) and (1f). 
In turn, this results in an elongation (shortening) of the $[111]$ ($[1\bar{1}1]$) bonds. 
Since the hopping terms in a tight-binding model are attributed to wave-function overlaps, we should then expect the terms corresponding to the $[111]$ ($[1\bar{1}1]$) bonds to increase (decrease) in magnitude. 
Indeed, for more sophisticated tight-binding models like sp$^3$d$^5$s$^*$, strain is also incorporated through modified bond lengths~\cite{Niquet2009}.
For a minimal tight-binding model like Eq. (\ref{HamTB}), this effect can be incorporated, to linear order in $\varepsilon_{xy}$, as a nearest-neighbor hopping term with alternating signs. 

To derive an effective-mass Hamiltonian from the tight-binding Hamiltonian $H_{\text{TB}}$, it is useful to transform into the momentum-space basis, defined as 
\begin{equation}
    \ket{k} = 
    \frac{1}{\sqrt{N}} 
    \sum_{m} e^{i k z_m} \ket{m}, \label{kState}
\end{equation}
where $N$ is the total number of sites in the 1D tight-binding chain, $z_m = m a_0/4$ is the $z$ coordinate of site $m$, and $k$ is the momentum restricted to the first Brillouin zone, $-4\pi/a_0 < k \leq 4 \pi/a_0$. 
(Note, here, the similarity to the unfolded Brillouin zone scheme described in the main text.)
Evaluating $H_{\text{TB}}$ in this basis yields
\begin{equation}
\begin{split}
    \mel{k}{H_\text{TB}}{k^\prime} =& 
    4\delta_{k}^{k^\prime} 
    \bigg[
    t_1 \sin^2 \left(\frac{k a_0}{8}\right)
    +t_2 \sin^2 \left(\frac{k a_0}{4}\right)
    \bigg] %\\
    + \widetilde{V}(k - k^\prime)
    -\delta_{|k - k^\prime|}^{4\pi/a_0}
    2i \varepsilon_{xy} t_{\varepsilon} 
    \sin(\frac{k a_0}{4})
\end{split}, \label{HTBMom}
\end{equation}
where the term in square brackets contains the band-structure dispersion shown in Fig. 1c, and $\widetilde{V}$ is the Fourier transform of the real-space potential,
\begin{equation}
    \widetilde{V}(k) = 
    \frac{1}{N}
    \sum_m 
    V_m e^{-i k z_m}.
\end{equation}
The shear-strain term in Eq.~(\ref{HTBMom}) contains the important selection rule $|k-k^\prime| = 4\pi/a_0$, which arises by Fourier transforming the $(-1)^{n}$ factor in the real-space representation of $H_{\text{TB}}$ in Eq. (\ref{HamTB}), and reflects the elongation and shortening of the $\left[111\right]$ and $\left[1\bar{1}1\right]$ bonds, respectively. 
This selection rule has the effect of coupling states exactly halfway across the %extended 
Brillouin zone. For example, as shown in Fig. 1\textbf{c} (green arrow) of the main text, shear strain couples the valley minimum at $k=-k_0$ to the virtually occupied state at $k=-k_0 + 4\pi/a_0$. 

Next, we decompose the momentum basis states into two classes ($\pm$) and relabel the states to allow for a useful decomposition of the Hamiltonian. Explicitly, we define
\begin{equation}
    \ket{k}_\pm = \ket{k \pm 2\pi/a_0},
\end{equation}
where the momentum is measured relative to $\pm 2\pi/a_0$.
We thus obtain the intra-class matrix elements 
\begin{equation}
\begin{split}
    \,_{\pm}\!\mel{k}{H_{\text{TB}}}{k^\prime}_\pm =& 
    2\delta_{k}^{k^\prime}\Big[ 
    -t_1 - t_2
    \mp t_1 \sin \left(ka_0/4\right) %\\
    - t_2 \cos \left(ka_0/2\right)\Big]
    +\widetilde{V}(k - k^\prime)
\end{split},  \label{IntraMtxElem}
\end{equation}
and the inter-class matrix elements
\begin{equation}
    \,_{+}\!\mel{k}{H_{\text{TB}}}{k^\prime}_{-} = 
    \widetilde{U}(k - k^\prime)
    - \delta_{k}^{k^\prime}2i \varepsilon_{xy} t_{\varepsilon} \cos \left(\frac{ka_0}{4}\right),  \label{InterMtxElem}
\end{equation}
where $\widetilde{U}(k)= \widetilde{V}(k +4\pi/a_0)$ is the potential energy, shifted in momentum space by $4\pi/a_0$. 
The low-energy eigenstates of $H_{\text{TB}}$ have large spectral weight from states $\ket{k}$ with momenta near the valley minima, $k \approx \pm k_0$. 
Since these minima occur near $k=2\pi/a_0$, we are justified in expanding the matrix elements in Eqs.~(\ref{IntraMtxElem}) and (\ref{InterMtxElem}) to $\mathcal{O}(k^2)$ near these points, yielding
\begin{align}
    \,_{\pm}\!\mel{k}{H_{\text{TB}}}{k^\prime}_\pm \approx& 
    ~\delta_{k}^{k^\prime}\Big[ 
    -2t_1 - 4t_2
    \mp t_1 k a_0/2 \label{IntraMtxElem2} %\\ 
     + t_2 k^2 a_0^2/4\Big]  
    +\widetilde{V}(k - k^\prime), \\ %\notag, \\
    \,_{+}\!\mel{k}{H_{\text{TB}}}{k^\prime}_{-} \approx& 
    -\delta_{k}^{k^\prime}2i \varepsilon_{xy} t_{\varepsilon}\left(1 - k^2 a_0^2/16\right) \label{InterMtxElem2} %\\
    +\widetilde{U}(k - k^\prime). %\notag 
\end{align}
We now note that Eqs.~(\ref{IntraMtxElem2}) and (\ref{InterMtxElem2}) are the matrix elements that would be obtained for a real-space, continuum representation of the Hamiltonian, by replacing $\delta_{k}^{k^\prime} k \rightarrow -i\partial_z$, and Fourier transforming $\widetilde{V}$ and $\widetilde{U}$ back into real space. 
The only difference between these discrete and continuum representations is that, technically, the tight-binding model is restricted to $\ket{k}_\pm$ with $-2\pi/a_0 < k \leq 2\pi/a_0$, while $k$ is unrestricted in the continuum model. 
In practice however, this difference is unimportant, because the low-energy eigenstates are dominated by basis states $\ket{k}_\pm$ with small $k$. 
We are therefore justified in using either tight-binding or continuum models to explore the physics of shear strain in this system. 
The continuum Hamiltonian is explicitly given by
\begin{equation}
    H_{\text{cont}} = 
    \begin{pmatrix}
        H_{++} & H_{+-} \\
        H_{+-}^\dagger & H_{--}
    \end{pmatrix},
\end{equation}
where 
\begin{align}
    H_{\pm \pm} &= -\frac{\hbar^2}{2 m_l} \partial_z^2 + V(z) \mp i\eta \partial_z, \\
    H_{+ - } &= V(z) e^{-i (4\pi/a_0) z} - 2i\varepsilon_{xy} t_{\varepsilon}\left(1 + (a_0^2/16)\partial_z^2\right).
\end{align}
Here, $m_l = 0.92 m_e$ is the effective mass characterizing the curvature at the bottom of the valley minima in Fig. 1\textbf{c}, and $\eta$ is the slope of the dispersion at $k = \pm 2\pi/a_0$.

Finally, we gauge away the linear $\partial_z$ term in the diagonal Hamiltonian components by expressing the wave function components as
\begin{equation}
    \psi_\pm(z) = e^{\mp ik_1 z} \psi^\prime_\pm (z), \label{psiTransformation}
\end{equation}
where $k_1 = \eta m_l/\hbar$. Plugging Eq.~(\ref{psiTransformation}) into the Schr\"{o}dinger equation for the Hamiltonian $H_{\text{cont}}$ yields a transformed Hamiltonian.
Under this transformation, we find $H_{++}=H_{--}$.
Identifying these terms with $H_0$ and the transformed $H_{+-}$ term with $H_v$, we finally obtain the effective-mass Hamiltonian, Eq.~(1) of the main text, which includes the effects of both shear strain and the WW. 
In this formulation, the parameter $C_1$ is related to the shear-strain hopping parameter $t_\varepsilon$, above, by $C_1 = 2 t_\varepsilon\left(1 - a_0^2 k_1^2/16\right)$.
Here, we adopt the values $t_{\varepsilon}=0.90$~eV and $C_1 = 1.73~\text{eV}$, such that the shear-strain dependent valley splitting, obtained from our minimal tight-binding model, closely matches results from the sp$^3$d$^5$s$^*$ model. 

%\section{Comparison of valley splitting in shear-strained quantum wells, with and without germanium concentration oscillations}
\section{Supplementary Note 2}

%%%%%%%%%%%%%%%%%%%%%%%%%%%%%%%%
%%%%%%%%%%%%%%%%%%%%%%%%%%%%
\begin{figure}[t]
\begin{center}
\includegraphics[width=.6\textwidth]{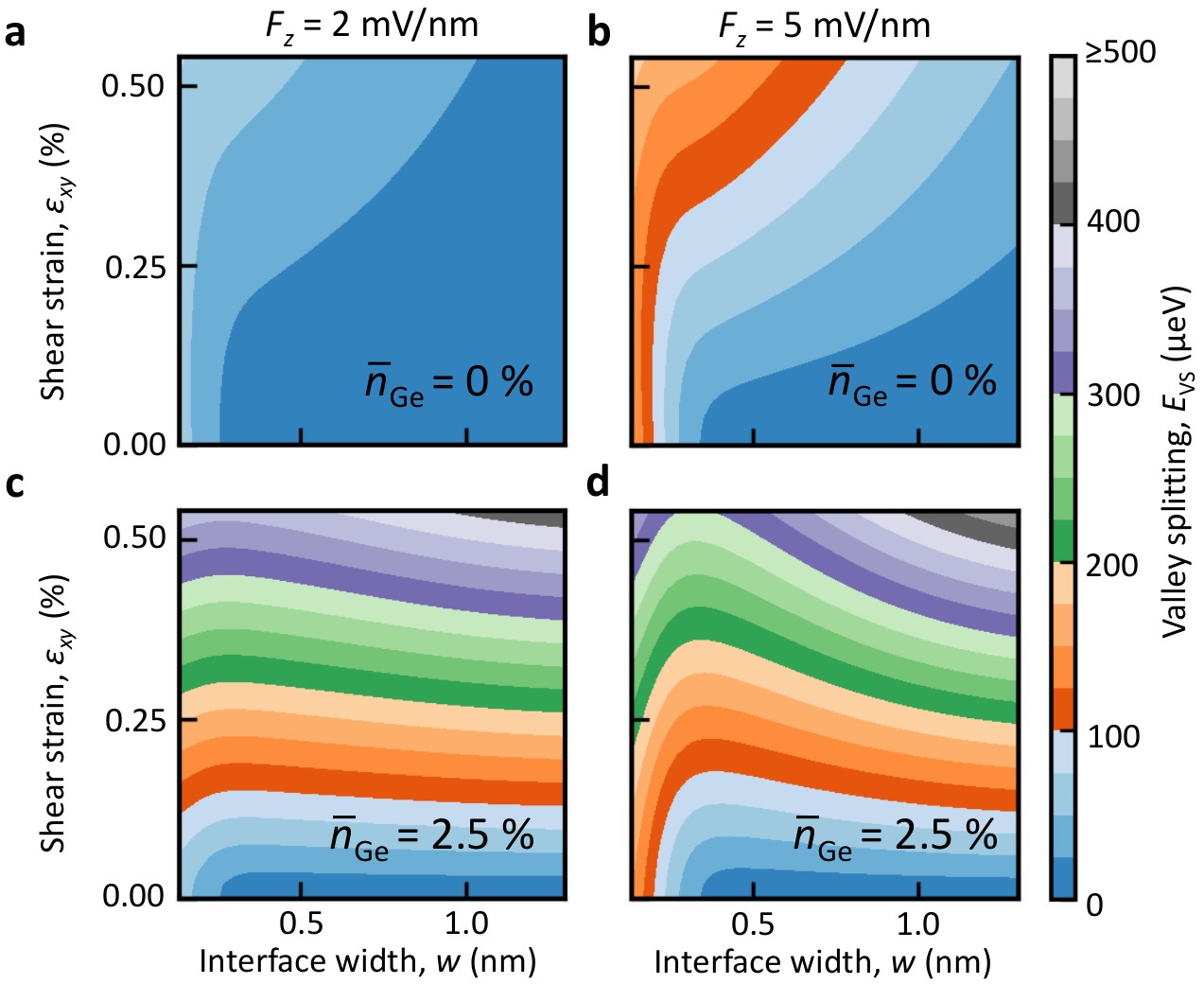}
\end{center}
\vspace{-0.5cm}
\caption{\textbf{Tight-binding simulations of valley splitting in a shear-strained quantum well, with and without Ge concentration oscillations.} \textbf{a}, \textbf{b}, Valley splitting $E_{\text{VS}}$ is computed as a function of shear strain $\varepsilon_{xy}$ and interface width $w$, in the absence of a WW ($\bar{n}_{\text{Ge}} = 0$). 
We consider electric fields of \textbf{a}, $F_z = 2$ or \textbf{b}, $5~\text{mV/nm}$. 
Here we see that a large $\varepsilon_{xy}$ is needed to reach the large-$E_{\text{VS}}$ regime, except in the limit of ultra-sharp interfaces, $w \lesssim 0.32~\text{nm}$.
We also find that the larger electric field (in \textbf{b}) yields a larger $E_{\text{VS}}$ compared to \textbf{a}, because the wave function is concentrated closer to the interface. 
Generally, the behavior in the main portion of these two panels is dominated by the third term of Eq.~(5), while the behavior in the bottom-left region is independent of $\varepsilon_{xy}$ and is dominated by the first term of Eq.~(5).
\textbf{c}, \textbf{d}, $E_{\text{VS}}$ is computed for STRAWW devices, with the ideal Ge oscillation period of $\lambda = 1.68~\text{nm}$, and an average Ge concentration of $\bar{n}_{Ge} = 2.5\%$ for the concentration oscillations in the quantum well. Again we consider electric fields of \textbf{c}, $F_z = 2$ or \textbf{d}, $5~\text{mV/nm}$.  
Compared to \textbf{a} and \textbf{b}, the valley splitting in \textbf{c} and \textbf{d} increases much more quickly with $\varepsilon_{xy}$, and to a greater degree. 
Additionally, $E_{\text{VS}}$ is nearly independent of the interface width and electric field $F_z$ for $w \gtrsim 0.5~\text{nm}$, which is outside the ultra-sharp interface regime, indicating that STRAWW is fundamentally not reliant on the presence of a sharp interface.
For these two panels, the behavior on the right-hand side is dominated by the fourth term of Eq.~(5), while the left-hand side is dominated by the first term of Eq.~(5), particularly in the limit of small $\varepsilon_{xy}$.}
\label{FIGS1}
\vspace{-1mm}
\end{figure}
%%%%%%%%%%%%%%%%%%%%%%	
%%%%%%%%%%%%%%%%%%%%%%%%%%%%%%%%

In the main text, we claimed that the valley splitting obtained in STRAWW geometries is preferable and more robust than comparable results in shear-strained Si/SiGe quantum wells without a WW. 
This statement is verified numerically in Supplementary Figure~\ref{FIGS1}, where we calculate  $E_{\text{VS}}$ from the sp$^3$d$^5$s$^*$ tight-binding model, as a function of shear strain $\varepsilon_{xy}$ and interface width $w$, in the absence (panels a and b) or presence (panels c and d) of a WW. We consider electric fields of $F_z = 2$ and $5~\text{mV/nm}$, and the Ge concentration profile $n_{\text{Ge}}$ defined in Methods. For context, recent experiments \cite{Wuetz2021} have measured interface widths of $w \approx 0.8~\text{nm}$ for Si/SiGe quantum wells when fitting to the Ge concentration profile $n_{\text{Ge}}$ defined in Methods.
We note that the enhancement of $E_{\text{VS}}$ due to shear strain, independent of Ge concentration oscillations, is captured in the third term of the inter-valley coupling matrix element $\Delta$, defined in Eq.~(5) of the main text. 
This effect has also been discussed in previous works~\cite{Ungersboeck2007,Sverdlov2008,Adelsberger2023}. 
However, it is clear from Eq.~(5) that the coupling term is only non-vanishing when the envelope function $|F_0(z)|^2$ has spectral weight at the wavevector $k=2k_1$.
A similar situation was discussed in Ref.~\cite{Losert2023}, in the context of the short-wavelength WW: in the absence of wiggles, $|F_0(z)|^2$ can have significant weight at large wavevectors only in the presence of an ultra-sharp interface, which produces wavevectors with a broad spectrum.
For example, in Supplementary Figure~\ref{FIGS1}, significant strain-induced enhancement of the valley splitting ($>100$~$\mu$eV) occurs only for very large shear strains ($\varepsilon>0.3$\%), with very sharp interfaces, corresponding to just a few atomic monolayers.

See the caption of Supplementary Figure~\ref{FIGS1} for further details and comments.

\end{document}